%% file: paper.tex
\documentclass[conference]{IEEEtran}
\PassOptionsToPackage{table}{xcolor}
\usepackage{letltxmacro}
\LetLtxMacro\oldttfamily\ttfamily
\DeclareRobustCommand{\ttfamily}{\oldttfamily\csname ttsize\endcsname}
\newcommand{\setttsize}[1]{\def\ttsize{#1}}%
\setttsize{\small}
\usepackage{booktabs}
\usepackage{adjustbox}
\usepackage{tabularx}
\usepackage{xspace}
\usepackage{enumitem}
\usepackage{url}
\usepackage{amsmath}
\usepackage{xcolor}
\usepackage{pifont}

\newcommand{\system}[0]{\textsc{MADEA}\xspace}
\newcounter{magicrownumbers}

\newcounter{examplecounter}
\newcommand\newexample{\addtocounter{examplecounter}{1}\theexamplecounter}
\newcommand{\cmark}{\ding{51}}
\newcommand{\xmark}{\ding{55}}

\newcommand{\parheading}[1]{\vspace{2pt}\noindent{}\textbf{{#1}}}
\newcommand{\eg}{e.g.,}
\newcommand{\ie}{i.e.,\xspace}
\newcommand{\etc}{etc.}

\newcommand{\dpidefinition}{deep packet inspection}
\newcommand{\dpi}{DPI}

\newcommand{\iot}{IoT\xspace}
\newcommand{\prv}{{\ensuremath{\sf{\mathcal Prv}}}\xspace}
\newcommand{\vrf}{{\ensuremath{\sf{\mathcal Vrf}}}\xspace}
\newcommand{\ra}{{\ensuremath{\sf{\mathcal RA}}}\xspace}
\newcommand{\ta}{{\ensuremath{\sf{\mathcal TA}}}\xspace}

\begin{document}
\title{\system: A \underline{Ma}lware \underline{De}tection \underline{A}rchitecture for IoT
blending Network Monitoring and Device Attestation}

\author{\IEEEauthorblockN{Renascence Tarafder Prapty}
\IEEEauthorblockA{University of California Irvine\\
rprapty@uci.edu}
\and
\IEEEauthorblockN{Rahmadi Trimananda}
\IEEEauthorblockA{University of California Irvine\\
rtrimana@uci.edu}
\and
\IEEEauthorblockN{Sashidhar Jakkamsetti}
\IEEEauthorblockA{Bosch Research\\
sashidhar.jakkamsetti@us.bosch.com}
\and
\IEEEauthorblockN{Gene Tsudik}
\IEEEauthorblockA{University of California Irvine\\
gts@ics.uci.edu}
\and
\IEEEauthorblockN{Athina Markopoulou}
\IEEEauthorblockA{University of California Irvine\\
athina@uci.edu}}

\maketitle
\input{content/abstract}
\input{content/introduction}
\input{content/background}
\input{content/system_model}

\input{content/design.tex}
\input{content/evaluation}
\input{content/discussion}
\input{content/conclusion}
\bibliographystyle{unsrt}
\bibliography{reference}
\input{content/appendix.tex}
\end{document}

%% file: content/abstract.tex
\begin{abstract}
Internet-of-Things (IoT) devices are vulnerable to malware and require new mitigation techniques due to their limited resources. To that end, previous research has used periodic Remote Attestation (\ra) or Traffic Analysis (\ta) to detect malware in IoT devices. However, \ra is expensive, and \ta only raises suspicion without confirming malware presence. To solve this, we design \system, the first system that blends \ra{} and \ta{} to offer a comprehensive approach to malware detection for the IoT ecosystem. \ta builds profiles of expected packet traces during benign operations of each device and then uses them to detect malware from network traffic in real-time. \ra confirms the presence or absence of malware on the device. \system{} achieves 100\% true positive rate. 
It also outperforms other approaches with 160$\times$ faster detection time. Finally, without \system{}, effective periodic \ra can consume at least $\sim$14$\times$ the amount of energy that a device needs in one hour.
\end{abstract}

%% file: content/introduction.tex
\section{Introduction}
Internet-of-Things (IoT), namely smart devices are becoming increasingly widespread. They are encountered in many settings,
including home, office, factory, public venues, and transportation of all sorts. Such devices are specialized sensors,
actuators, or hybrids thereof; and their computational and communication resources are usually meager due to size, cost,
and power constraints. Unsurprisingly, they represent an attractive set of targets for malware and other exploits.
For example, the most infamous IoT malware incident was the Mirai botnet attack that infected hundreds of thousands of 
DVRs, IP cameras, routers, and printers in 2016~\cite{antonakakis2017understanding}. There are many other types
of IoT-focused malware exploiting diverse and often new attack vectors. Malware's job is made easier in terms
of infection scalability by the general consumer tendency towards monocultures, as witnessed by the immense popularity 
of certain ``smart'' voice assistants, doorbells, cameras, and appliances. This motivates the need for new malware mitigation
techniques geared for the IoT ecosystem, distinct from those used to combat more traditional malware~\cite{alrawi2021circle}. 

\parheading{Limitations of Prior Work.}
The research community has been keenly aware of the IoT security issues and many prior works proposed various mitigation
techniques. A large body of such work is focused on Remote Attestation (\ra{}) -- a means of measuring the internal 
software state of a remote device to determine whether it is infected~\cite{sancus,grisafi2022pistis,ammar2020simple,vrasedp}. 
Most of prior work proposed and evaluated \textit{new} \ra{} techniques but often side-stepped an important practical
aspect: how to schedule the \ra? Doing it infrequently makes instantaneous detection of malware impossible. In the best case, it increases the detection latency and, in the worst case, allows the malware to come and go between successive measurements and thus escape undetected. Meanwhile, doing it too often is impractical and wasteful since each \ra instance consumes resources by 
taking the attesting device away from its main task(s), as well as expending its power due to computation and communication.

Another major direction of prior work is detection of malicious activity via Traffic Analysis (\ta)~\cite{tekiner2022lightweight,marin2019deep,meidan2018n,alrashdi2019ad,wozniak2020recurrent}.
Most \ta{} approaches rely on Machine Learning (ML) techniques, which are typically not lightweight 
and use many traffic features. 
Packets sent and received by \iot{} devices are captured and their relevant features are extracted and used to build data samples for training ML models. Then, during the monitoring phase, classifiers label each new data 
sample as either benign or suspicious. 
Nevertheless, \ta, by itself, only flags suspicious
activity and does nothing to confirm whether a device is indeed infected. Note that some \ta techniques blacklist
suspected devices without confirming infection and discard all of their traffic, which is overkill.

\parheading{\system{} Approach.}
To address the above issues, this paper constructs \system{}, a system that blends \ra and \ta, resulting in a more comprehensive IoT malware detection than either of the two alone. To the best of our knowledge, \system is the first to leverage the benefits of both \ta and \ra. We provide a comparison of \system with prominent existing approaches of \ta \cite{nguyen2019diot,meidan2018n} and \ra \cite{vrasedp} in Table \ref{tab:novelty table}.
\begin{table}[t!]
	\centering
        \caption{Comparison of \system with prior approaches.}
	\begin{tabularx}{\linewidth}
{p{15mm} p{20mm}p{18mm} p{18mm}}
  \toprule
  \textbf{Approach} & \textbf{Search Vector} & \textbf{Instantaneous} & \textbf{Malware} \\
  & & \textbf{Detection} & \textbf{Confirmation} \\
  \midrule
  \system{} & Network Packet \& Device Memory & \cellcolor{green!25}\cmark & \cellcolor{green!25}\cmark\\
  \midrule
  N-BaIoT  & Network Packet & \cellcolor{green!25}\cmark & \cellcolor{red!25}\xmark \\
  \midrule
  D\"IoT  & Network Packet & \cellcolor{green!25}\cmark & \cellcolor{red!25}\xmark \\
  \midrule
  VRASED  & Device Memory & \cellcolor{red!25}\xmark & \cellcolor{green!25}\cmark \\
  \bottomrule
\end{tabularx}
\vspace{-1em}
\label{tab:novelty table}
\end{table}

\system{} is aimed at a typical smart home/office 
IoT device deployment. Conveniently, 
traffic patterns of such devices tend to be limited and predictable. They usually communicate with only a handful of external endpoints and, as shown in prior work~\cite{trimananda2020packet,oconnor2019homesnitch}, variations in packet sizes are minor. 
Infected devices show diverging traffic patterns by communicating with different IP addresses and using different packet sizes, typically talking to their Command-and-Control Center (CCC) or other infected devices.

\system{} consists of three main components: \textit{Profiler},  \textit{Monitor}, and \textit{Attester}. 
Profiler trains one \ta{} profile per device by collecting all packets generated by the device during a 
typical (benign) operation, and extracting and storing three features per packet: endpoint addresses, 
packet length, and direction.
Next, Monitor detects potentially malicious activity in real time through \textit{whitelisting}: 
a packet is considered as suspicious if it does not match the features of any packets stored in the corresponding device profile.
Whenever that occurs, \system{} invokes Attester on the suspected device to determine whether it is infected. Infection can be detected by changes in the program memory of the device.
If an infection is confirmed, \system{} reports this incident; otherwise, it includes the triggered traffic pattern in the benign patterns database. Our experiments demonstrate that \system{} achieves 100\% True Positive Rate (TPR) with at most 1.2\%
false positive rate (FPR), on average.

\parheading{Contributions.} 
First, \system{} performs \textit{lightweight \ta} of IoT devices via {\em per-packet} detection, using only three features, to match each packet in real-time against device profiles while maintaining high detection accuracy. Notably, \system outperforms machine learning based state of the art approach, \cite{nguyen2019diot} with a 1.05$\times$ better TPR and 160$\times$ faster detection time. 

Second, \system{} benefits from triggering \ra only when a suspicious traffic pattern is detected. 
This also allows periodic/scheduled \ra{} to be done infrequently, which avoids unnecessary \ra{} 
instances that would otherwise waste device resources and take devices away from their actual primary 
tasks that might be safety-critical. In our evaluation, without the \ra{}-\ta{} combination, a camera with 1.84 W power consumption would need to consume \textit{an additional $\sim$26 Wh of power in a year for periodic attestation without any guarantee of detecting the malware.}

Furthermore, the entire \system{} system is made available as an open-source implementation at \cite{madea-repo}. Its proof-of-concept includes two components: 
(1) a prototype IoT device based on
Raspberry Pi 4, in which the Attester module is implemented
by extending the Raspbian OS to measure the currently
running processes and (2) an emulated router, 
also based on Raspberry Pi 4, which profiles and monitors device traffic.

\parheading{Organization.} 
Section~\ref{sec:background} provides background and related work in \ra{} and \ta{} for IoT malware.
Next, Section~\ref{sec:systemmodel} describes \system{}'s system and adversary models and
Section~\ref{sec:design} lays out the high-level design and implementation details.
Then, Section~\ref{sec: Evaluation} presents the evaluation results, followed by
Section~\ref{sec:discussion}, which discusses various aspects and limitations of \system.
Section~\ref{sec:conclusion} concludes the paper.

%% file: content/background.tex
\section{Background \& Related Work} \label{sec:background}
This section provides some background on targeted devices, \ra, and \ta, and on their interaction.
\subsection{Targeted Devices} \label{subsec:targeted_devices}
This work focuses on resource-constrained IoT devices specifically designed for sensing and actuation tasks, some of which perform safety-critical functions (e.g., fire/smoke detectors, smart locks, glass vibration alarms, and surveillance/motion cameras). They are equipped with micro-controller units (MCUs) such as ARM Cortex-M series or mid-range MCUs like ARM Cortex-A series. To support security features, some MCUs provide extended hardware features like memory protection units (MPUs) or memory management units (MMUs).

This work is applicable to a wide range of IoT devices equipped with some form of root of trust. However, very simple devices  (so-called ``bare metal'') that lack security features are considered out-of-scope.

\noindent {\bf Network Connectivity.}
IoT devices are often connected to the Internet and other peer devices, either directly or via an intermediary, 
such as a controller hub or a router. This paper focuses on devices that use routers for connectivity.
Such devices are equipped with at least one network interface, such as WiFi, Bluetooth, cellular, wired Ethernet, or Zigbee.
We focus on WiFi traffic since WiFi is the most common network interface present on consumer IoT devices~\cite{blevswifi}. 
Albeit, any other network media (wired or wireless) can be easily supported by \system.

\subsection{Remote Attestation: \ra}
\ra{} is a security service that enables a trusted party (\vrf) to measure current software state of an untrusted remote 
device (\prv). \ra protocol is usually realized as follows:
\begin{itemize}
    \item \vrf sends an \ra request to \prv that contains a challenge.
    \item \prv receives the request, computes a secure measurement of its current software state and sends it to \vrf.
    \item \vrf verifies the result and decides whether \prv is in a valid/healthy state.
\end{itemize}
The measurement is obtained by: (1) hashing (using a suitable cryptographic hash function, such as SHA-256) the software, and 
then (2) computing either a Message Authentication Code (MAC) or a digital signature over the hash digest and the challenge. 
Computing a MAC requires \prv to share a key with \vrf, while computing a signature requires \prv to have a 
private key with the corresponding public key known to \vrf. Both approaches require \prv to have some form of secure storage 
for the (shared or private) key. The measurement is typically computed 
within a Trusted Computing Base (TCB) that relies on trusted hardware and/or software, depending on the type of \ra: software-based,
hardware-based, or hybrid. 
Hardware-based \ra~\cite{tpm,sancus,kil2009remote,MQY10} relies on dedicated secure
hardware components, e.g., TPM~\cite{tpm} or Intel SGX~\cite{sgx}). 
However, such hardware is normally too costly (in multiple aspects) for most IoT devices.
Software-based \ra~\cite{KeJa03,seshadri2004swatt,ammar2020simple,grisafi2022pistis,gligor} requires no secure 
hardware features. However, it assumes an ideal environment with reliable communication even in the presence of a remote attacker, which is difficult to achieve in real world scenarios.

Hybrid \ra~\cite{smart,vrasedp,tytan,trustlite} claims security equivalent to hardware-based \ra, at a lower cost. Such methods require minimal hardware support and shift the bulk of complexity to the (trusted) software component. However, hybrid \ra still requires hardware modification and, therefore, is not readily applicable to off-the-shelf IoT devices.

For our proof of concept implementation, we use a software-based \ra{} method suitable for off-the-shelf consumer IoT devices.  
However, we expect that \system{} could be used in tandem with many \ra{} methods. Section \ref{sec:discussion} discusses the compatibility of \system with other \ra{} techniques.

\subsection{Traffic Analysis: \ta}\label{subsec:network-monitoring}
In \ta, a specific software/hardware combination (device $x$) is profiled in a controlled setting to identify its traffic characteristics under normal operating conditions. These characteristics (also referred to as $x$'s traffic fingerprint or profile) are then used to identify $x$ on any network by matching live traffic patterns to $x$'s fingerprint. Network fingerprints can include bandwidth, timing of TCP connections, port numbers, IP addresses, packet sizes, protocol identifiers, directions, etc. For example, 
\cite{trimananda2020packet} develops packet-level event signatures of IoT devices using only sequence numbers and the
size and direction of TCP packets.  

Some early \ta\ techniques relied on \dpidefinition{} (\dpi{})~\cite{moore2005,ma2006}.
Surprisingly, \dpi{} also initially proved effective for smartphones~\cite{miskovic2015} and \iot{}~\cite{feng2018}, 
perhaps due to lax security practices in the early days of these technologies.
However, gradually traffic encryption became more common~\cite{razaghpanah2017,alrawi2019,huang2020}. Consequently, new fingerprinting techniques were proposed that identify individual functionality on IoT devices even
if the application-layer payload is encrypted, \eg{}~\cite{oconnor2019,trimananda2020packet,acar2020,varmarken2022fingerprintv}. 
Network fingerprinting has various useful applications such as intrusion detection, traffic prioritization, and user activity profiling. However, attackers also use it to reconnoiter a network for an effective attack.

Although they vary greatly in terms of complexity and collected traffic features, these techniques focus only on a subset 
of \iot{} traffic that uniquely identifies a specific device or a certain event occurring on it. In contrast, \system{} monitor component
operates as a whitelist. Therefore, it needs to learn the complete (normal) traffic profile of an \iot{} device, which it achieves by
blending a simple passive \ta technique (to flag suspicious patterns) with \ra (to perform infection checking), 
as described in Section \ref{sec:design}).

Another body of prior work focused on detecting malicious activity via \ta\ based on Machine Learning (ML) 
techniques.~\cite{tekiner2022lightweight,marin2019deep,nguyen2019diot,alrashdi2019ad,wozniak2020recurrent}.
These techniques tend to use numerous traffic features, resulting in heavy-weight ML models.
Typically, packets to/from \iot{} devices are captured and relevant features are extracted in order to build data samples. 
During the training phase, ML models are trained with labeled benign and malicious data samples. 
During the monitoring phase, classifiers label each data sample as either benign or suspicious. 
A related approach involves ML models trained with only benign data~\cite{meidan2018n,rey2022federated}.
In contrast, \system{} uses very few traffic features which results in a very lightweight, heuristic-based detection model.

\subsection{\ra-\ta\ Synergy}
One important practical aspect of \ra{} mostly avoided by prior work is: {\em exactly when to perform it?} 
Two intuitive choices are to perform \ra{} at: (1) scheduled, e.g., at fixed and/or random intervals,  
and/or (2) on-demand, e.g., before using \prv. Both approaches have pros and cons. Though easy to configure, the former 
has two disadvantages: First, it leaves potentially
long time gaps between successive \ra\ instances which allows malware to come and go in the interim and remain undetected.
Second, it is generally quite wasteful due to \ra's resource consumption on \prv. If the device performs safety-critical tasks,
performing \ra can divert it from such tasks or at least slow it down.

Meanwhile, most \ta\ techniques do not go beyond detection of suspicious activity, \ie{} they do not inspect the device to confirm that it is actually infected, though some involve suspected device quarantine. This is clearly insufficient for comprehensive IoT malware mitigation. Even quarantine represents potential over-reaction since the suspected device might in fact not be infected. The only way to make sure is to inspect the suspected device, \eg{} via \ra. 
\system{} uses lightweight \ta to flag suspicious IoT traffic patterns and triggers \ra\ for all suspected devices. 
If an \ra\ result reflects malware presence, the user (\ie{} administrator/owner) is immediately notified.
Otherwise, in case of a false positive (no malware detected based on the \ra\ result), \system{} ``learns'' 
by whitelisting the observed traffic pattern.

%% file: content/system_model.tex
\section{System \& Adversary Models} \label{sec:systemmodel}
This section describes \system\ system model and adversarial assumptions. 

\subsection{System Model} 
\system considers four entities: an IoT device, a gateway, a user device, and a remote server.
\begin{figure}
    \centering
    \includegraphics[width=\linewidth]{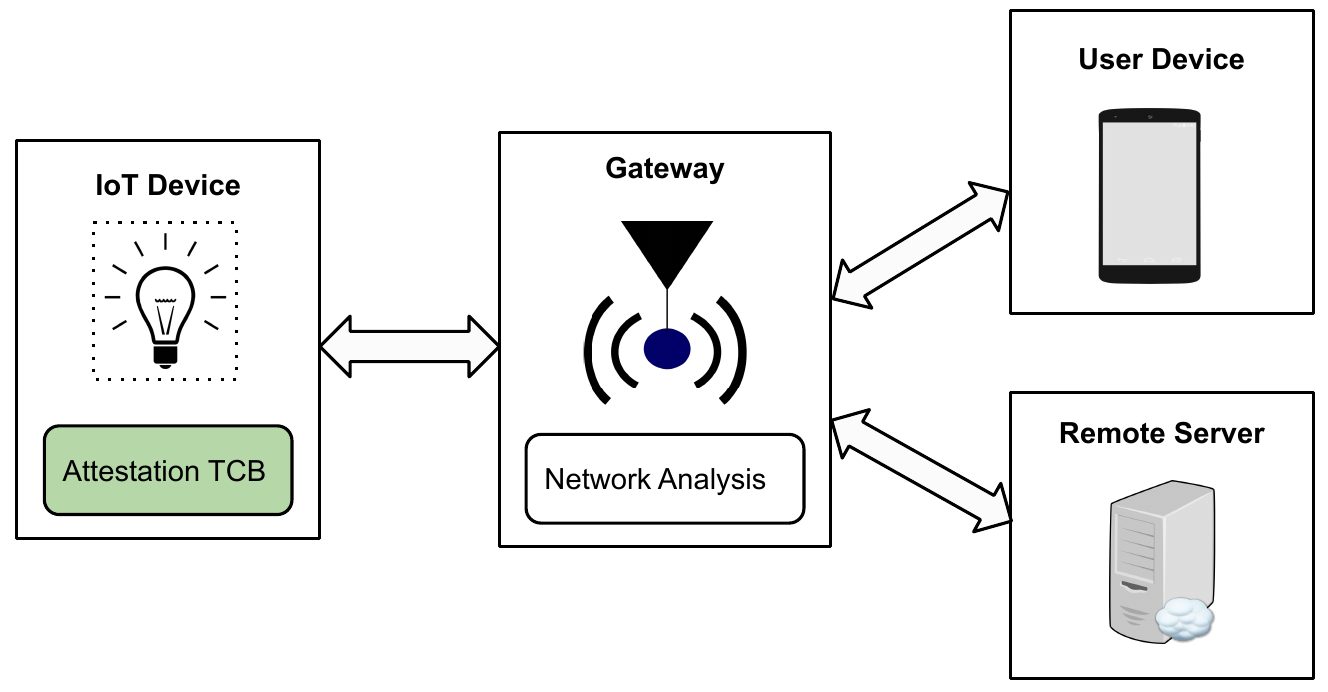}
    \caption{\system system model.}
    \label{fig:system_model}
\end{figure}

IoT device is a typical smart device (as in Section \ref{subsec:targeted_devices}) that is the main target of the adversary. It is assumed to be installed in a private space (such as a home/office setting) or a public place (such as a grocery store or a hotel), connected to a gateway. A gateway can be a local router, an IoT controller hub, or any other gateway that connects a set of IoT devices directly to (or towards) the Internet. We assume that an IoT device is equipped with a Trusted Computing Base (TCB), that performs device attestation, as in Section \ref{sec:background}. A remote server is a remote back-end server (generally hosted in the cloud) maintained by the IoT service provider. The server is responsible for registering, updating, and assisting the IoT device. A user device is a personal item (such as a smartphone, smartwatch, laptop, or tablet) that controls the IoT device via a dedicated app or a web interface. 

\system{} works in an IoT deployment with at least one IoT device, one or more user devices, and one gateway, as shown in Figure \ref{fig:system_model}. User devices communicate with IoT devices either directly or via the server. All packets from or to IoT devices pass through the gateway: we assume that peer-to-peer communication is not allowed. The gateway captures packets and makes decisions based on their features, without keeping the state of on-going traffic flows/connections. In the context of \ra, IoT devices and gateway play the roles of \prv and \vrf, respectively.

\subsection{Adversary Model} \label{subsec:advesarymodel}
We make the following assumptions about adversarial capabilities: 
\begin{itemize}
    \item \textit{Remote adversary:} infects IoT devices with malware over the network without having physical access to them, for example, 
    by exploiting a software vulnerability in an application running on IoT devices or similar to Mirai, through telnet by exploiting default/weak telnet credentials.
    \item \textit{TCB}: IoT device's TCB is trusted and not subject to compromise. Each TCB has a distinct pair of public and secret keys for computing signatures as part of \ra. Each public key is known to the gateway.
    \item \textit{Malware-generated traffic:} We assume that malware must generate some amount of traffic, in order to 
    communicate with its Command-and-Control Center (CCC). In most cases, malware activities are coordinated by CCC -- a remote server maintained by the adversary which controls a network of infected devices. This network is usually called a botnet and infected devices are called zombies/bots. CCC sends various commands to be executed by IoT devices, \eg{} to download code, infect other devices, or launch attacks. 
    \item \textit{Trusted entities:} We assume that the gateway, server, and user devices are sufficiently powerful computing platforms that can protect themselves from malware and remain trusted. Moreover, since the service provider is trusted, each newly deployed IoT device is assumed to be initially healthy. DNS resolver and other routing protocols are considered to be uncompromised: an adversary cannot spoof IP address of CCC to match the IP address of the remote server.
\end{itemize}

\subsubsection{DoS and Other Attacks}
An adversary can launch a DoS attack
on an IoT device by flooding it with a high volume of traffic from
the outside, i.e., from beyond the gateway. Alternatively, the adversary can use
infected IoT devices to launch a DoS attack on other peer (healthy) devices. \system detects both DoS attacks. Since we assume
that the gateway runs on a trusted high-end platform, DoS attacks on it are considered out-of-scope. DoS attacks on IoT devices are further discussed in Section VI.

We consider all physical attacks (i.e., those requiring adversary's
physical presence) to be out of the scope of MADEA. The
same holds for malware that does not involve any of its own
incoming or outgoing traffic. (Malware that targets actuation
capabilities and acts autonomously would fall into this category.)
Section \ref{sec:discussion} discusses this in more detail.

%% file: content/design.tex
\section{Design \& Implementation} \label{sec:design}
\begin{figure}
    \centering
    \includegraphics[width=\linewidth]{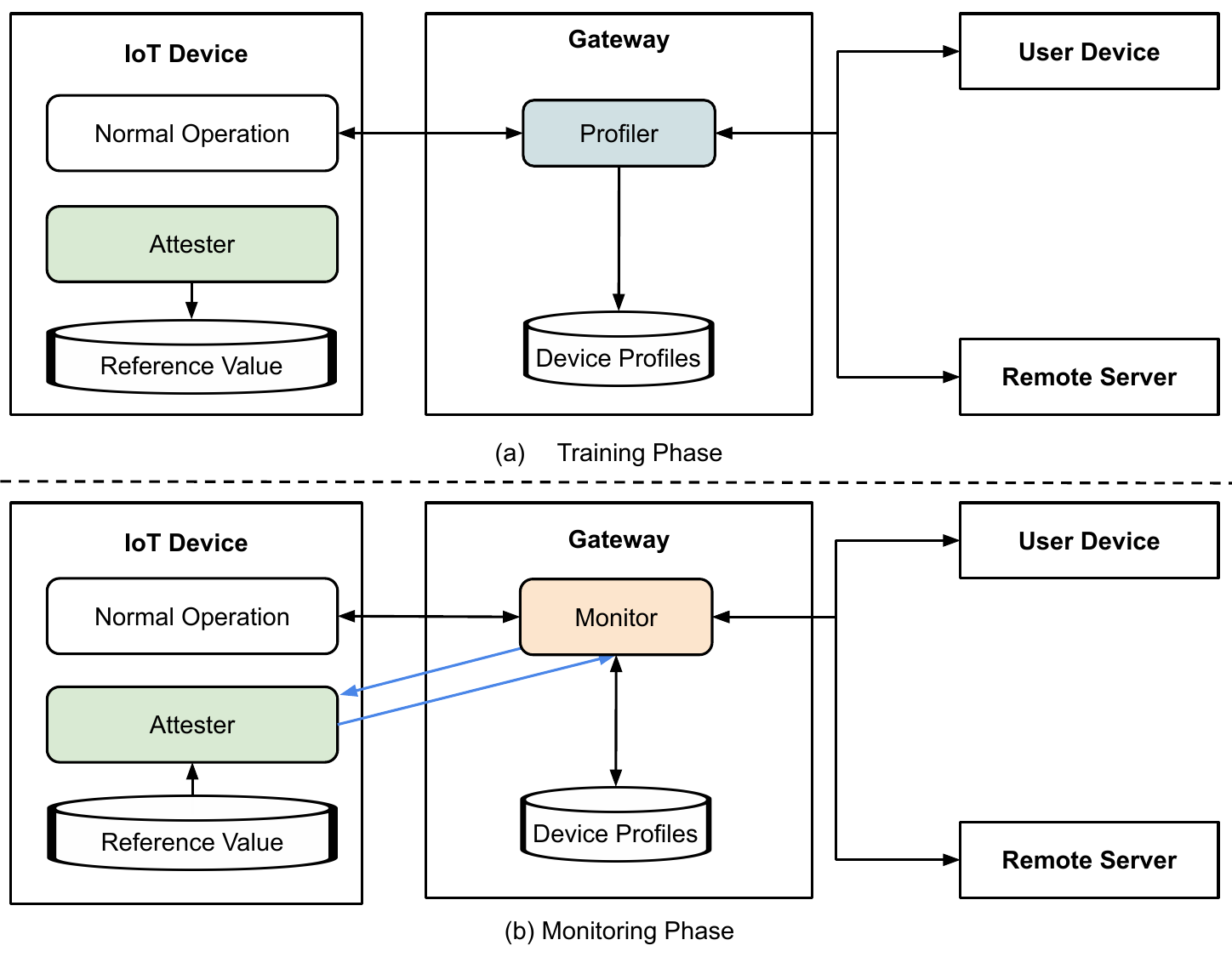}
    \caption{\system overview: blue arrows represent the feedback loop between Attester and Monitor.
    }
    \label{fig: system overview}
\end{figure}

This section discusses the main components of \system{}: Profiler, Monitor, and Attester (see Figure~\ref{fig: system overview}). 

\subsection{Profiler}\label{subsec:profiler}
The first step of \system{} is building a device profile for each IoT device during the training period when each device is newly deployed and assumed to be malware-free. This task is performed by Profiler running on the gateway. All normal events of an IoT device (\eg{} turning on/off as well as brightness and color control in case of a smart bulb) are triggered manually using the corresponding controller app in the user device. User manuals and options in controller apps are consulted to develop the list of all possible events for the IoT device. This list is easy to build since \system{} targets simple to moderately complex devices such as smart bulb, smart plug, robot vacuum cleaner, camera, \etc{} Profiler can build profiles for multiple IoT devices simultaneously, and the training is performed online.

Profiler refreshes its state after every 20,000 packets to avoid stalling caused by large PCAP files due to limited RAM and processing power. It stores IoT device profiles in a database shared with Monitor, using a hash map for fast access. The MAC addresses of devices to be trained are also stored in the database and read by Profiler at the beginning of each iteration, allowing for dynamic changes in the list in each iteration.

\begin{figure}[t]
    \centering
    \includegraphics[width=\columnwidth]{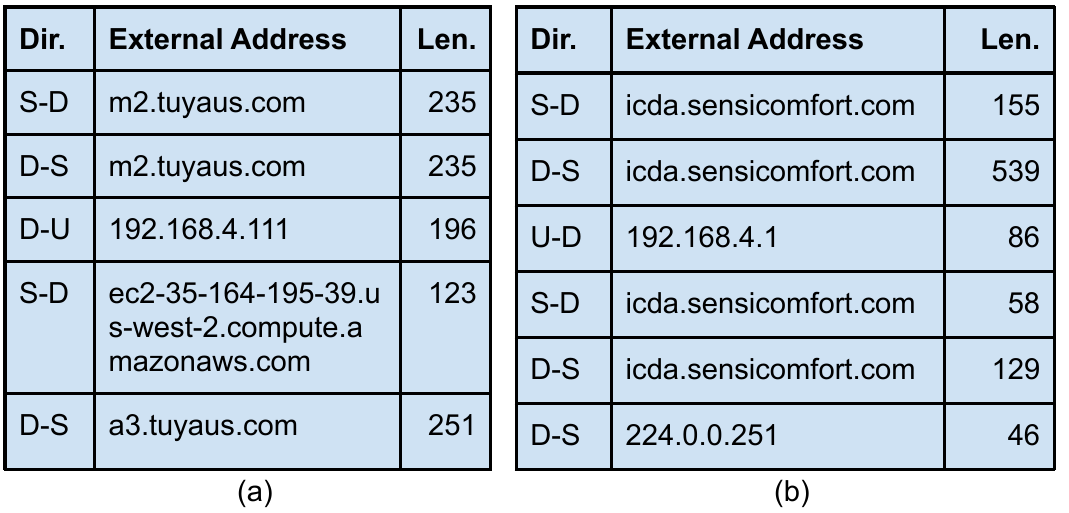}
    \caption{Partial profiles of: \textbf{(a)} Lumiman Bulb, \textbf{(b)} Sensi Thermostat (S-D=\textit{Server-to-Device}, D-S=\textit{Device-to-Server}, D-U=\textit{Device-to-User}, U-D=\textit{User-to-Device}). 
    Note: See Appendix A for full profile of Sensi Thermostat. 
    For more examples, see \cite{MADEA_dataset}.} 
    \label{fig: profile example}
\end{figure}

\parheading{Device Profile.}
A device profile refers to the profile of the expected network traffic to/from the IoT device. Figure \ref{fig: profile example} shows example profiles for two commodity IoT devices: Lumiman bulb \cite{lumimanbulb}, and Sensi thermostat \cite{sensithermostat}. Each profile can be viewed as a table, where each row represents a unique group of packets called profile entry. Thus, each profile consists of a list of profile entries. Each profile entry consists of the device MAC address, external address, packet direction and length. 

External address is the IP address belonging to the entity that exchanges packets with the IoT device. It refers to either its corresponding server or user device. This is the most important feature since malware-generated packets usually have different IP addresses. This holds as long as malware does not compromise DNS and/or the routing protocol. 

Legitimate packets sent to/received from a specific external address by an IoT device usually have slight length variations.  
It is worth mentioning that \system{} can effectively detect malware packets without this feature. However, we include it because packet length provides additional information while being very lightweight.  

Packet direction reflects whether the IoT device is the source or destination of the packet, \ie{}
\begin{itemize}
    \item \textit{Server-to-Device}: sent by the server to the IoT device.
    \item \textit{Device-to-Server}: sent by the IoT device to the server.
    \item \textit{Device-to-User}: sent from the IoT device to the user device in the same network.
    \item \textit{User-to-Device}: sent from the user device to the IoT device within the same network.
\end{itemize}
Note that we choose four different values for packet direction to make profile entry more intuitive and informative for human users. Just two directions (to/from the IoT device) would suffice for \system{} to work properly. Moreover, \system{} does not require users to be in the same LAN as the IoT device. A remote user device communicates with the IoT device through the server and is treated the same as communications between the IoT device and server.

Packet direction plays an important role since suspicious packets generated by the IoT device might cause more concern than those sent to the IoT device. For example, packets sent to the IoT device may indicate malware payload being downloaded or malicious command. In contrast, packets sent from the IoT device might indicate exfiltration of sensed data or malware being sent to another IoT device. Moreover, used together, packet direction and length make the packet matching criteria stricter. Oftentimes, packets of a specific length are exchanged only in one direction and not the other.  

Other features that could be taken into account include the packet sequence, frequency, protocol, or packet volume per unit of time. We do not include them in \system{} because, as we found out, the set of three aforementioned features suffices to detect malware with high accuracy (see Section \ref{sec: Evaluation}). Furthermore, this makes the packet matching algorithm in Monitor lightweight and efficient.

Packets with the same external address, direction, and length are grouped together. This grouping is independent of packet sequencing and packets in the same group can be captured at any time during the training phase. We assume that each profile contains entries for all possible groups of packets that can be transmitted to and from the IoT device. 

\parheading{Building Profile.}
Figure~\ref{fig: system overview} shows how Profiler builds a profile. 

PyShark~\cite{pyshark} library captures and processes packets from the IoT device via its LiveCapture interface. A BPF filter, configured with the MAC address of the IoT device, filters captured traffic. 

To construct a profile entry, five fields are extracted from each packet: source IP address, destination IP address, source MAC address, destination MAC address, and packet length. Further, \system{} assumes that the IoT device is connected to the LAN defined by the gateway. Thus, either the source or destination MAC address must be the actual MAC address of the IoT device, whereas the other MAC address is the MAC address of the gateway.  

\parheading{Configuration Packets.}
Configuration packets of standard protocols, such as NTP, DHCP, DNS, MDNS, BOOTP, ARP, EAPOL, are not included in the profile. Such packets are typically transient and do not provide meaningful insight into the recurrent nature of IoT device traffic. However, \system{} considers both TCP and UDP packets for the profile. Many prior network fingerprinting mechanisms \cite{trimananda2020packet,oconnor2019homesnitch} consider only TCP packets. The fact that many modern IoT devices use UDP makes \system{} more inclusive and robust.

\parheading{Determining Direction \& External Address.}
Packets are associated with a specific IoT device using its MAC address. This way, if the IoT device changes its IP address, its profile is not affected. MAC addresses are used to determine directions and IP addresses. If the list of MAC addresses in the database contains the source MAC address of the packet, then the destination IP address is considered the external address. In that case, the direction is \textit{Device-to-Server}. On the other hand, the source IP address becomes the external address if the destination MAC address is in the list of MAC addresses and the direction is \textit{Server-to-Device}. Finally, if both source and destination IP addresses are local addresses, the direction is either \textit{Device-to-User} or \textit{User-to-Device}. 

\parheading{Mapping IP Addresses to Hostnames.}
In our experiments with numerous devices, we observed that IP addresses of servers can change frequently, leading to unstable profiles. Hostnames, on the other hand, remain unchanged or change rarely. Therefore, using hostnames as external addresses is preferable. However, this requires an IP-to-hostname mapping in \system{}, as hostnames are resolved to IP addresses through DNS packets,
and the associated IP addresses are used in subsequent communication.

\system{} builds IP-address-to-hostname mappings using DNS response packets instead of creating profile entries for captured DNS packets. It extracts hostnames from the \textit{Query} field and resolved IP addresses from the \textit{Answer} field to create a mapping from each IP address to its corresponding hostname.

If source/destination IP address is not a local address, \system{} tries to find the corresponding hostname from the IP-address-to-hostname mapping. If this fails, \system{} performs reverse DNS lookup using the IP address. If a hostname is returned, \system{} uses it; otherwise, it uses the IP address. 

\parheading{Example \newexample.} 
In this example, we use profiles of Lumiman Bulb and Sensi Thermostat from Figure~\ref{fig: profile example}.

\noindent \textit{Lumiman Bulb.}
In the first profile entry, the bulb is the destination of the packet. Thus, the direction is \textit{Server-to-Device} and the source address is the external address. The source IP address is resolved to a hostname (\ie{} \textit{m2.tuyaus.com}) using IP-address-to-hostname mapping and that hostname denotes the external address. 
During the development of \system{}, the hostname \textit{m2.tuyaus.com} resolves to 21 different IP addresses, \eg{} \textit{54.212.163.173}. Interestingly, apart from Lumiman Bulb, another IoT device (\ie{} HBN plug) also exchanges packets with the same external address. Thus, these two devices are likely to be associated with the same manufacturer and talk with the same servers. The second profile entry has the same external address and length as the first entry, but the opposite direction. 
In the third profile entry, both source and destination IP addresses are local and the bulb is the source. Thus, the direction is \textit{Device-to-User} and the IP address of the user device becomes the external address. In the fourth profile entry, the external address is a hostname resolved through reverse DNS lookup (\ie{} the source IP address cannot be resolved using IP-address-to-hostname mapping) and the direction is \textit{Server-to-Device}. The original source IP address is 35.164.195.39 and the hostname implies that the server is hosted on AWS cloud. 

\noindent\textit{Sensi Thermostat.}
In the Sensi Thermostat profile in Figure \ref{fig: profile example}(b), the external address of the last profile entry is an IP address and the direction is \textit{Device-to-Server}. It means that the destination IP address cannot be resolved even through reverse DNS lookup.

\subsection{Monitor}
Monitor attempts to match each packet to/from the IoT device to some profile entry. We assume that all normal events have been reflected in the profile during the training phase. Hence, if no match is found for a packet, \system{} assumes it to be suspicious (\ie{} potentially generated by malware). Appendix B provides more insight into the differences between observed malware-generated and normal packets. One concern about this approach is that updates in either the IoT device or server can change the benign traffic pattern. Please refer to Section \ref{sec:discussion} for details about how \system{} handles updates. Any device can be added to the list of IoT devices to be monitored as Monitor runs constantly. Similar to Profiler, it refreshes its state after every 20,000 packets. 

\parheading{Matching a Packet.} 
During the monitoring phase, Monitor constructs an entry for every packet following the same approach that Profiler performs ( see Section~\ref{subsec:profiler}). Then, Monitor checks for a matching profile entry based on MAC address, external address, packet direction and length. If no exact match is found, it performs partial string matching for external address with at least 80\% similarity while ensuring the top-level domain name is the same. 
 
For instance, \textit{ec2-35-164-195-39.us-west-2.compute.amazonaws.com} will match with \textit{ec2-13-235-158-83.ap-south-1.compute.amazonaws.com} since both partially match and the top-level domain is \textit{amazonaws.com}. 80\% similarity is chosen based on empirical observation. This percentage gives a negligible number of false positive packets for hostnames found through reverse DNS lookup while keeping the level of similarity significant. We define false positive as a benign packet being detected as suspicious by \system{}. With 85\% and 90\% similarity matching, we observe more false positives. Partial string matching is considered because the external address changes slightly periodically in some IoT devices, \eg{} Ring doorbell and Blink camera. MAC address, and packet direction and length are still matched exactly. 

\parheading{Example \newexample.}
In this example, let us consider a packet captured by Monitor. The following five properties are extracted from it: 
\begin{enumerate}
    \item source MAC address: \textit{64:16:66:49:3e:cb}.
    \item destination MAC address: \textit{dc:a6:32:ce:31:63}
    \item source IP address: \textit{192.168.4.230}
    \item destination IP address: \textit{35.186.98.64}
    \item packet length: \textit{284}
\end{enumerate}

\system{} maintains a list of MAC addresses of the monitored devices in the database. It also maintains a hash map of all profile entries. Monitor observes that the list contains source MAC address. Thus, Monitor determines that the device MAC address is \textit{64:16:66:49:3e:cb} and the packet direction is \textit{Device-to-Server}. Furthermore, the destination IP address is translated to the hostname \textit{oculus9353-us1.dropcam.com} that becomes the external address. 
Monitor forms an entry as a tuple (MAC address: \textit{64:16:66:49:3e:cb}, external address: \textit{oculus9353-us1.dropcam.com}, packet direction: \textit{Device-to-Server}, packet length: \textit{284}) and finds a matching profile entry from the hashmap. 
If a matching profile entry is found for Monitor's entry, it means that such a packet has been seen by Profiler during the training phase and, thus, the packet is considered legitimate. The Monitor's entry may fail to match any profile entry because of two reasons: (1) there is no profile entry that can be matched with Monitor's entry when Monitor inspects the MAC address, external address, and packet direction; or (2) there are profile entries that can be matched with Monitor's entry when Monitor inspects MAC address, external address, and packet direction; however, the packet length in profile entry does not match the packet length in Monitor's entry.
   
We have observed that mismatched packet lengths contribute the most to FPR. However, the security impact of mismatched packet lengths is relatively low, as long as external address is matched. Thus, \system{} offers relaxed matching, in which Monitor excludes packet length when matching Monitor's entry to a profile entry.

Finally, in the case that Monitor detects a suspicious packet, \ie{} a packet whose entry does not match with any profile entry, it then calls a program called Attester to verify the state of the IoT device.

\subsection{Attester}
Attester runs locally on the IoT device and it is responsible for attesting the software state of the IoT device when requested. It works on the assumption that the software state of the IoT device is deterministic, measurable, and unchanged unless the IoT device is infected by malware. It is to be noted that the software state can also change due to legitimate software updates. Section~\ref{sec:discussion} discusses how this case is handled by \system{}.
 
During the training phase, the software state of a healthy IoT device is measured as the \textit{reference value} that indicates that the IoT device is not compromised. During the monitoring phase Attester measures the software state when called by Monitor. Monitor's call includes a new challenge every time. The measured value is compared with the reference value locally by Attester as sending the full measurement result to Monitor can be expensive in terms of network bandwidth. 
If the values match, Attester reports an acknowledgment indicating that the IoT device is healthy. Otherwise, it reports that the IoT device has been infected by malware. In this case Attester's report includes the divergent measurement result for further analysis.
The Attester's report is always signed using the IoT device's secret key, and the signature includes the challenge so that the gateway can authenticate that the report is fresh and indeed generated by the IoT device.

In \system{} proof of concept implementation, Attester is a Python program running on a Raspberry Pi emulated smart bulb. Traditional device attestation that involves measuring the full program memory is not suitable for RPi's virtual memory which is managed by OS. We can't determine the physical addresses of processes in memory, and, thus, cannot determine which memory region to attest. Additionally, measuring the large RAM (in the scale of gigabytes) of a RPi would be slow and impractical. Therefore, Attester attests the binaries of the currently running processes on the RPi instead. 

Attester works on the assumption (backed by observation 4, see \ref{subsec:effectiveness-feedback-loop}) that the processes running on the IoT device, at any given time, are deterministic and limited in number. Hence, it is possible to maintain a list of expected processes and hash values of their corresponding binaries. If any IoT device is infected by malware, a new malware process would be running on it or the malware would embed itself in an existing process which would change the hash value of the corresponding binary file. Attester inspects this hash value change to detect malware on the IoT device. This attestation is supported by extensive experiments conducted on the emulated bulb for evaluation. Section \ref{sec: Evaluation} provides more details on these experiments.

During the training period, Attester builds a list of expected processes and corresponding hash values for the IoT device. 
When called by Monitor, Attester obtains a list of all currently running processes by calling the Linux command \texttt{ps aux | less > [output file]} which directs the output to a text file. Next, the output file is parsed to extract the PID of each process. PID is used to obtain the location of the binary file for each process. This location is in the format of \texttt{/proc/[PID]/exe}. Finally, \system{} calculates the hash value of the binary file of each process using the command \texttt{sha256sum /proc/[PID]/exe}.  

This list is then compared against the list of expected processes. If there is any new process in the current process list or an expected process with a wrong hash value, the IoT device will be considered compromised (\ie{} infected by malware) and Attester reports this to Monitor: Attester will also send the list of the new processes and/or the processes with the wrong hash values. 
Otherwise, if there is no new process and all current processes have the same hash values as calculated during training, Attester will send the attestation report back to Monitor indicating that the IoT device is healthy.

\begin{figure}
    \centering
    \includegraphics[width=\columnwidth]{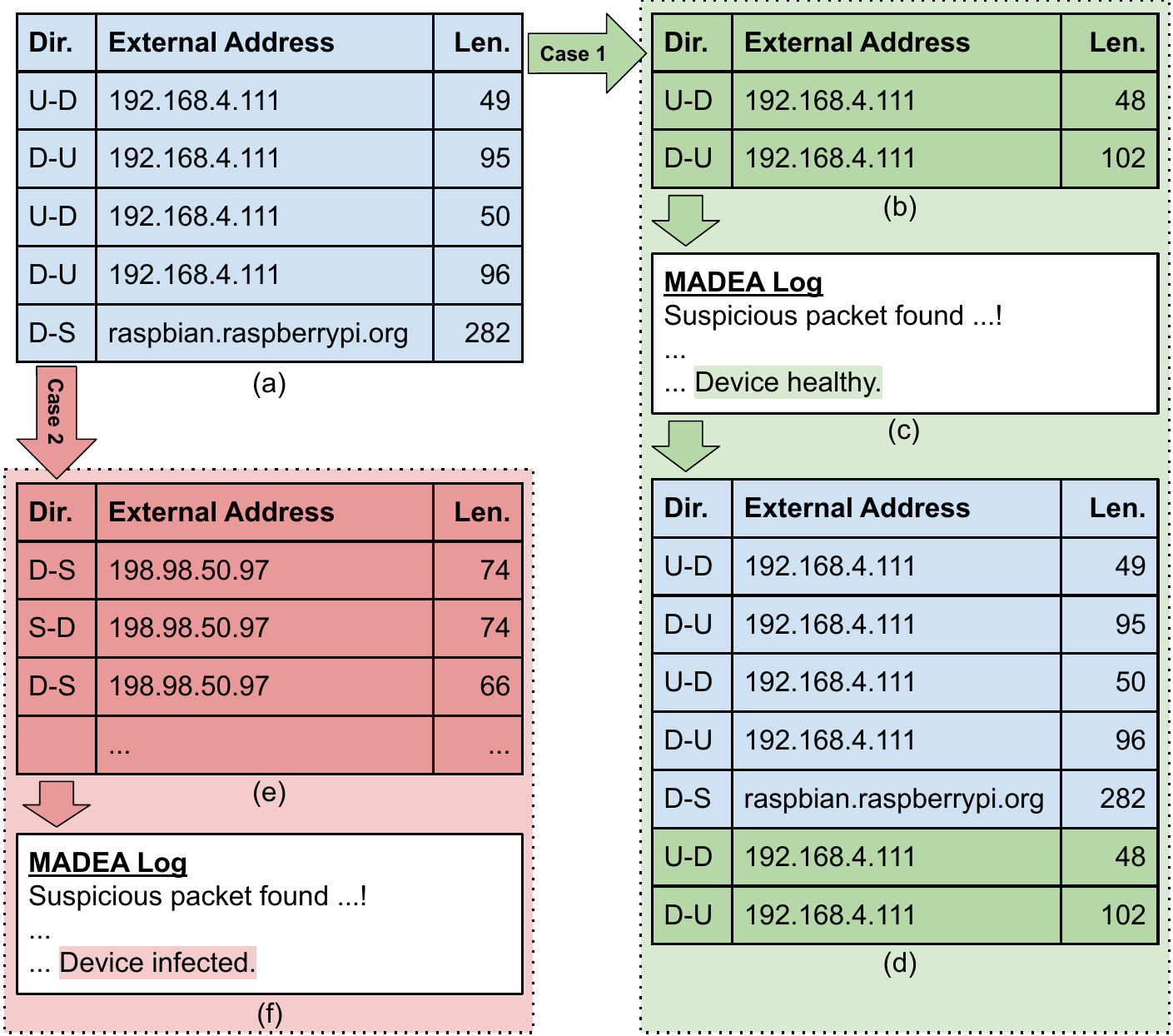}
    \caption{\textbf{Case 1}---Malware not detected by Attester:
    (a) the RPi smart bulb's profile, (b) packets for a new command (\ie{} status message), (c) Monitor's log, and (d) the RPi smart bulb's profile updated by Monitor. \textbf{Case 2}---Malware detected by Attester: (a) the RPi smart bulb's profile, (b) packets generated by malware, and (c) Monitor's log.}
    \label{fig: positive feedback}
\end{figure}

\subsection{Feedback Loop}\label{subsec:feedback-loop}
Monitor and Attester complement each other and create a feedback loop, which is a key contribution of \system{}.
During monitoring, if no matching profile entry is found for Monitor's entry of a particular packet, Monitor calls Attester to attest the IoT device. If Attester finds no malware (Case 1), it reports to Monitor that the IoT device is healthy and Monitor considers the triggering packet as a false positive. Next, it updates the IoT device's profile by creating a corresponding profile entry and appending it to profile to avoid future false positives prompted by similar packets. If Attester finds malware on the IoT device (Case 2), it reports to Monitor that the IoT device has been infected.  

Section~\ref{sec: Evaluation} provides an example of the feedback loop operation; Figure~\ref{fig: positive feedback} illustrates the process. Currently, Monitor logs the report message for further analysis by device owner. Ideally, Monitor would send the report directly to the owner as an alert for Case 2.

%% file: content/evaluation.tex
\section{Evaluation}\label{sec: Evaluation}
We next present the evaluation for \system{}.  
Section~\ref{subsec:experimental-setup} outlines the experimental setup for \system{}.
Section~\ref{subsec:monitoring-devices} presents the evaluation of \system{} in monitoring devices in three datasets.
Section~\ref{subsec:effectiveness-feedback-loop} presents a case study to illustrate the effectiveness of \system{}'s feedback loop, which is a main contribution in this paper.
Section~\ref{subsec:performance-overheads} characterizes \system{}'s performance overheads.
Section~\ref{subsec: comparison-prior-works} compares the performance of \system{} with three prior works. Finally, Section~\ref{subsec: energy-savings} illustrates the energy savings over periodic attestation due to the feedback loop.
\subsection{Experimental Setup}\label{subsec:experimental-setup}
Our experimental setup follows the \system{} system model and architecture shown in Figures~\ref{fig:system_model} and~\ref{fig: system overview} respectively. We use a Rasberry Pi 4 (RPi4)~\cite{raspberry_pi4} machine as the gateway, configured as a WiFi access point. It is based on quad-core Cortex-A72 (\ie{} ARMv8-M architecture) 64-bit Systems-on-Chip (SoC). The board runs at 1.5GHz with 8GB LPDDR4-2400 SDRAM. Additionally, it is equipped with 2.4GHz and 5.0GHz IEEE 802.11B/g/n/ac WLAN, Bluetooth 5.0, double-true Gigabit Ethernet, and a 32GB SD card as storage. Thus, with these specifications, RPi4 is comparable with commodity home routers. Profiler and Monitor are implemented as Python programs which run on the gateway.

\parheading{Sandbox for Malware.}
The gateway is connected to a NETGEAR 4G LTE Broadband Modem (LM1200)~\cite{netgear_modem} with Ethernet cable to access the Internet. NETGEAR Broadband Modem uses a T-Mobile SIM card for network connectivity. With this, we create a sandbox for safety reasons since we experiment with malware. Connecting the gateway directly to our institutional LAN would potentially make other LAN devices vulnerable to malware infection. 

\parheading{Smart Home Setup.}
We use a smart home setup to evaluate \system{}. However, \system{} can be implemented at scale to suit the deployment in a commercial setting. Our setup consists of 13 off-the-shelf devices and one RPi4 emulated smart bulb as the IoT device. 
Table~\ref{tab:device list} provides details about the devices. 

Off-the-shelf devices are chosen based on their popularity and ratings on Amazon~\cite{amazon}. We also consider a wide range of functionality: from plugs to cameras and vacuum-cleaning robots (\ie{} iRobot Roomba). The RPi smart bulb is implemented on a RPi4 with 2GB RAM. 
The RPi bulb is controlled by a Python program. This program can accept requests and send back responses over UDP connections. It has two main parts: bulb controller and attester. The bulb controller receives commands from the user device over the internet and turns on/off a LED based on the received command. Attester receives attestation request from the gateway, performs attestation, and sends back attestation result.

Additionally, we implement three malware CCC (Mirai, Bashlite, Lizkebab) on a Linux laptop with an Intel(R) Core(TM) i7-8550U CPU running at 1.80GHz and equipped with 8GB RAM. This Linux laptop is also connected to the gateway. Alternatively, malware CCC can be hosted outside the network, but we choose to host it inside our sandbox network due to safety issues.

We first collect network traffic (in the training phase) from the 13 devices to build \system{} dataset and, in turn, the device profiles without introducing any malware infection. 
Table~\ref{tab:device false positive} shows the numbers of packets from the traffic collection and profile entries for each device profile.

\begin{table}[t!]
        \scriptsize
	\centering
 \caption{Smart home devices used in \system{} evaluation.}
\begin{tabularx}{\linewidth}{p{26mm}p{48mm}} 
  \toprule
   \textbf{Device Name} & \textbf{Details} \\ 
  \midrule
   Amazon Smart Plug & Amazon Smart Plug \\ 
   HBN Smart Plug Mini & HBN Smart Plug Mini 15A \\ 
   Ring Doorbell & Ring Doorbell with camera (2020 release)\\ 
   Blink Mini Camera & Blink Mini compact indoor plug-in smart security camera \\ 
 Nest Camera & Nest indoor security camera \\ 
   Lumiman Smart Bulb & Lumiman multi-colored and warm-to-cool white LED light bulb\\ 
     Kasa Smart Bulb & Kasa Smart Bulb KL110 white light bulb \\ 
     LIFX Smart Bulb & LIFX A19 multi-colored light bulb \\ 
   ULTRALOQ U-Bolt Pro & ULTRALOQ U-Bolt Pro smart lock with Bluetooth and keypad\\ 
   Sensi Thermostat & Emerson Sensi smart thermostat \\ 
   Nest Protect Smoke Alarm & Nest Protect smoke alarm, smoke detector, and carbon monoxide detector\\ 
   Rachio Sprinkler & Rachio Sprinkler third generation with 8-zone sprinkler controller \\ 
   iRobot Roomba & iRobot Roomba 690 \\
  \midrule
     RPi Smart Bulb & Emulated smart bulb on Raspberry Pi 4 board \\ 
  \bottomrule
\end{tabularx}
\vspace{-1em}
\label{tab:device list}
\end{table}

\begin{table}[t!]
	\scriptsize
	\centering
        \caption{Experimental results of monitoring uninfected devices in smart home setup. The last row gives the \textbf{Total} for \textbf{\# Training Packets (Trng. Pkts.)}, \textbf{\# Monitoring Packets (Mnt. Pkts.)} and \textbf{\# Profile Entries}, and \textbf{Average} for \textbf{FPR}.}
	\begin{tabularx}{\linewidth}
{p{15mm} r r r r r}
  \toprule
  \textbf{Device Name} & \textbf{\# Trng.} &\textbf{\# Mnt.} & \textbf{\# Profile} & \multicolumn{2}{c}{\textbf{\underline{False Positive Rate (FPR)}}} \\
  & \textbf{Pkts.} & \textbf{Pkts.} & \textbf{Entries} & \textbf{End. \& Len.} & \textbf{End.}\\
  \midrule
  Amazon Plug & 8,317 & 11,017 & 119 & 0.87\% & 0.64\% \\ 
  HBN Plug Mini & 22,096 & 24,688 & 81 & 0.01\% & 0.00\% \\ 
  Ring Doorbell & 143,925 & 91,576 & 1590 & 0.39\% & 0.01\% \\ 
  Blink Mini Cam. & 52,839 & 1,477,004 & 132 & 0.00\% & 0.00\% \\ 
  Nest Cam. & 283,798 & 645,873 & 734 & 0.02\% & 0.02\% \\ 
  Lumiman Bulb & 42,274 & 49,855 & 113 & 0.07\% & 0.00\% \\ 
  Kasa Bulb & 8,667 & 6,434 & 146 & 1.09\% & 0.93\% \\ 
  LIFX Bulb & 1,207 & 408 & 21 & 0.00\% & 0.00\% \\ 
  ULTRALOQ U-Bolt Pro & 7,487 & 55541 & 137 & 0.05\% & 0.03\% \\ 
  Sensi Thermostat & 1,492 & 796 & 52 & 1.13\% & 0.00\% \\ 
  Nest Protect Smoke Alarm & 806 & 89 & 59 & 1.12\% & 1.12\% \\ 
  Rachio Sprinkler & 3,508 & 4,614 & 112 & 4.76\% & 0.11\% \\ 
  iRobot Roomba & 8,821 & 6,756 & 207 & 0.65\% & 0.25\% \\
  \midrule
  \textbf{Total/Average} & 585,239 & 2,374,651 & 3,503 & 0.78\% & 0.24\% \\
  \bottomrule
\end{tabularx}
\vspace{-1em}
\label{tab:device false positive}
\end{table}

\subsection{Monitoring Devices}\label{subsec:monitoring-devices}
\parheading{Monitoring Healthy Devices.}
First, we use the \system{} dataset to evaluate the performance of \system{}. In particular, we evaluate the performance of Monitor in monitoring uninfected devices using the created device profiles. 
Table \ref{tab:device false positive} provides FPR for each device. We measure FPR in two scenarios: (1) matching external address, packet direction, and length (see \textbf{End. \& Len.}); and (2) matching external address and packet direction (see \textbf{End.}) in Monitor's entry with the corresponding features in any profile entry in the device profile. \system{} incurs low FPR for all devices: 0.78\% and 0.24\% on average for both scenarios respectively.

\parheading{Public Dataset 1: BehavIoT.}
We also evaluate \system using a public dataset called BehavIoT~\cite{hu2023behaviot}. The BehavIoT dataset contains 32 uninfected IoT devices, among which three devices (\ie{} Amazon Plug, Ring Doorbell, and Nest Thermostat) are present in our setup as well. This dataset contains both LAN and WAN traffic for all possible events of each device. We only use the LAN traffic in alignment with our setup. 
We divide the LAN traffic per device per event into two equal parts. One part is used by Profiler to build device profiles (\ie{} training phase). The other part is used to evaluate the performance of Monitor (\ie{} monitoring phase). 

\begin{table}[t!]
\scriptsize
\centering
\caption{False Positive Rate of uninfected devices from BehavIoT dataset. The last row gives the \textbf{Total} for \textbf{\# Training Packets (Trng. Pkts.)}, \textbf{\# Monitoring Packets (Mnt. Pkts.)} and \textbf{\# Profile Entries (Prof. Ents.)}, and \textbf{Average} for \textbf{FPR}.}
\begin{tabularx}{\linewidth}{p{15mm} r r r r r}
  \toprule
  \textbf{Device Name} & \textbf{\# Trng.} &\textbf{\# Mnt.} & \textbf{\# Prof.} & \multicolumn{2}{c}{\textbf{\underline{False Positive Rate (FPR)}}} \\
  & \textbf{Pkts.} & \textbf{Pkts.} & \textbf{Ents.} & \textbf{End. \& Len.} & \textbf{End.}\\
  \midrule
 Tp-Link Bulb & 14,928 & 15,026 & 94 & 0.07\% & 0.00\% \\ 
Tp-Link Plug & 3,467 & 3,474 & 51 & 0.14\% & 0.00\% \\ 
Magic Home Strip Light & 2,592 & 2,597 & 19 & 0.00\% & 0.00\% \\ 
Smart Life Bulb & 5,599 & 5,511 & 23 & 0.18\% & 0.00\% \\ 
Nest Thermostat & 8,197 & 7662 & 245 & 1.59\% & 0.00\% \\ 
Bulb1 & 1,860 & 1,830 & 16 & 0.05\% & 0.00\% \\ 
Govee LED & 3,600 & 3,600 & 12 & 0.36\% & 0.36\%  \\ 
Gosund Bulb & 3,587 & 3,627 & 21 & 0.03\% & 0.00\% \\ 
Meross Door Opener & 621 & 647 & 21 & 2.94\% & 0.00\% \\ 
Amazon Plug & 1,297 & 1,263 & 41 & 0.39\% & 0.00\% \\ 
Echo Spot & 17,822 & 15,827 & 618 & 3.28\% & 1.24\% \\ 
Echo Show & 17,127 & 16,700 & 575 & 3.03\% & 0.25\% \\ 
Echo Dot & 3,803 & 4,247 & 131 & 2.33\% & 0.47\% \\ 
Philips Hub & 17,936 & 17,793 & 254 & 0.35\% & 0.00\%\\ 
SmartThings Hub & 1,568 & 1,498 & 81 & 0.47\% & 0.00\%\\ 
Aqara Hub & 862 & 801 & 40 & 0.49\% & 0.00\% \\ 
IKEA Hub & 2,170 & 2,172 & 51 & 0.00\% & 0.00\% \\
Wemo Plug & 11,390 & 10,795 & 87 & 0.17\% & 0.00\% \\ 
Wansview Cam. & 290,372 & 293,693 & 1104 & 0.05\% & 0.00\% \\ 
Lefun Cam. & 105,723 & 105,362 & 1,717 & 0.36\% & 0.00\%\\ 
Dlink Cam. & 229,960 & 236,918 & 1,576 & 0.04\% & 0.00\%\\ 
Microseven Cam. & 290,639 & 298,642 & 1,668 & 0.13\% & 0.00\%\\ 
Ring Door Bell & 238,173 & 234,583 & 3241 & 3.10\% & 0.00\% \\ 
Ring Cam. & 311,528 & 301520 & 12387 & 3.82\% & 0.00\% \\ 
Ubell Doorbell & 237,400 & 230,272 & 2581 & 0.06\% & 0.00\% \\ 
Wyze Cam. & 478,060 & 506,694 & 1,146 & 0.39\% & 0.00\% \\ 
Yi Cam. & 189,899 & 189,395 & 1,365 & 0.19\% & 0.00\% \\
Tuya Cam. & 477,039 & 497,072 & 293 & 0.01\% & 0.00\% \\
ICsee Doorbell & 268,628 & 271,611 & 1635 & 0.06\% & 0.00\% \\ 
SwitchBot Hub & 728 & 722 & 11 & 0.00\% & 0.00\% \\ 
iKettle & 18 & 33 & 6 & 6.06\% & 0.00\% \\
Google Home Mini & 8,089 & 7,900 & 861 & 9.29\% & 0.51\% \\
\midrule
\textbf{Total/Average} & 3,244,673
& 3,289,487 & 31,971 & 1.23\% & 0.09\% \\
\bottomrule
\end{tabularx}
\vspace{-1em}
\label{tab:device false positive behavIot}
\end{table}

Table \ref{tab:device false positive behavIot} provides the number of packets, profile entries, and FPR of each healthy device from this dataset.
\system{} shows relatively high FPR for a few devices from this dataset when the external address, packet direction, and length are matched. However, FPR is zero or very low when only the external address and packet direction are matched. Most of these devices are comparatively complex devices (\eg{} cameras and voice assistants). We observe that exact length matching is less effective for complex devices that may generate a more diverse traffic pattern compared to simpler devices (\eg{} plugs and bulbs). Instead, if we relax the matching criteria (\eg{} allowing a slight variation of 2 bytes in packet length for Google Home Mini), FPR is almost totally eliminated.

\parheading{Monitoring Infected Devices.}
We use the RPi smart bulb for this purpose. In the training phase, we build the device profile for the RPi smart bulb without infecting it with malware. Next, we infect the RPi smart bulb with Mirai, Lizkebab, Bashlite, and 100 randomly selected ARM malware binaries from~\cite{alrawi2021circle}. Monitor is able to detect suspicious packets with \textbf{100\% TPR} for all malware in the following aspects, namely: (1) the network payload sent from the malware CCC during bot installation, (2) the periodic heart-beat communication between the bot and the CCC, and (3) the attack commands sent from the CCC to the bot.

\parheading{Public Dataset 2: Cryptojacking.}
Finally, we evaluate \system{} using another public dataset that we call the Cryptojacking dataset \cite{tekiner2022lightweight}. The Cryptojacking dataset consists of both benign and IoT cryptojacking malware traffic collected from a Raspberry Pi, a laptop, a LG Smart TV, and a server. We exclude server data from our evaluation since it is not an IoT device.  
For each of these devices, \system{} successfully detects malware traffic with \textbf{100\% TPR}. 

Thus, our evaluation of \system{} on the three datasets and two scenarios (\ie{} healthy and infected devices) indicates that \system is effective in profiling and monitoring IoT devices.

\subsection{Effectiveness of Feedback Loop}\label{subsec:effectiveness-feedback-loop}
We use the RPi smart bulb to evaluate the feedback loop's effectiveness.
During the training phase, the bulb receives \textit{turn on, turn off} commands and sends responses. Entries for the corresponding packets generated by the two commands are included in the device profile of the RPi smart bulb. In the monitoring phase, we create two scenarios: \textbf{(1)} \textit{healthy bulb (Case 1)}: the user device sends a new command to the RPi smart bulb. The RPi smart bulb is uninfected at this stage, \textbf{(2)} \textit{infected bulb (Case 2)}: the RPi smart bulb becomes infected with malware and malware tries to communicate with its CCC.

\parheading{Healthy Bulb (Case 1).}
During the training phase the commands and responses exchanged between the user device and RPi smart bulb, when turning on and off the RPi smart bulb are:  
\begin{itemize}
    \item \textbf{Command}: turn\_on (packet length=49) \\ \textbf{response}: \{``status":``200",``message": ``Light bulb turned on."\}  (packet length=95)
    \item \textbf{Command}: turn\_off (packet length=50) \\ \textbf{response}: \{``status": ``200", ``message": ``Light bulb turned off."\}  (packet length=96)
\end{itemize}
From Figure \ref{fig: positive feedback} we can see that the profile of the RPi smart bulb includes profile entries for these packets. However, during the monitoring phase, another previously unseen command (\ie{} status message) is exchanged:
\begin{itemize}
    \item \textbf{Command}: status (packet length=48) \\ \textbf{response}: \{``status": ``200", ``message": ``Light bulb is currently off."\} \\ (packet length=102)\}
\end{itemize}

Monitor detects the command packet as suspicious and requests attestation from Attester inside the RPi smart bulb. Attester confirms that the RPi bulb is healthy since it does not find any new process or any expected process with a wrong hash value. As a result, Monitor adds the command packet to the bulb's profile. The same process is repeated for the response packet. Afterward, subsequent \texttt{status} command and response packets are treated as benign and not subject to further attestation.

\parheading{Infected Bulb (Case 2).}
The RPi smart bulb is intentionally infected with a malware binary. This binary is randomly selected from the malware repository of \cite{alrawi2021circle}.  Figure \ref{fig: positive feedback} shows some packets exchanged between the malware CCC and the zombified RPi smart bulb. All of those packets are confirmed as suspicious through the feedback loop. However, we illustrate the process for the first packet only.

Figure \ref{fig: positive feedback} shows that external address of the packet is \textit{198.98.50.97}, the packet length is 74 and direction is \textit{Device-to-Server}. However, no profile entry in the RPi smart bulb's profile matches those values. Thus, Monitor detects the packet as suspicious and calls Attester. Attester reports that new processes are running on the RPi smart bulb, which confirms that the RPi smart bulb is infected by malware, as shown in the warning message in the log file. 

\subsection{Performance Overheads}\label{subsec:performance-overheads}
The \ta{} technique deployed by \system{} to detect anomaly in the network traffic is very lightweight. The fourth column in Tables~\ref{tab:device false positive} and~\ref{tab:device false positive behavIot} denotes the number of profile entries, \ie{} maximum search space for matching packets of a specific device.
We observe that FPR is consistently low and TPR is consistently high in experimental settings and public datasets. 

\parheading{Runtime Overhead.}
In addition, the detection is fast. The total time for \system{} to extract endpoints and length properties from a packet, classify the packet as benign or suspicious, and verify the detection through the feedback loop is on average 1.6 ms. 
Note that this is an upper bound since our proof of concept implementation is in Python. An implementation in low-level languages such as C/C++ would result in better timing performance. 

\parheading{Storage Overhead.}
We can also determine the absolute upper bound of the total number of device profile entries stored in the memory of the router during the monitoring phase: $n=d * e * l $,
where $n$ represents the number of profile entries of all devices being monitored, $d$ represents the number of devices being monitored, $e$ represents the average number of packet direction-external address pairs for each device, and $l$ represents the average number of unique packet lengths per packet direction-external address pair. 

The average value of $l$ is 23 for the \system{} dataset and 58 for the BehavIoT dataset. The average value of $e$ is 13 for the \system{} dataset and 18 for the BehavIoT dataset.

\subsection{Comparison with Prior Works}
\label{subsec: comparison-prior-works}
In this section, we compare the performance of \system{} with two \ta-based  \cite{meidan2018n,nguyen2019diot} and one \ra-based \cite{vrasedp} systems from prior work. Table~\ref{tab:performance comparison} presents the details. \system{}'s TPR is 100\%, which means that \system can detect every malware-generated packet correctly. In other words, \system{} is more robust than D\"IoT~\cite{nguyen2019diot} and the same as N-BaIoT~\cite{meidan2018n}, VRASED~\cite{vrasedp}. The attack/malware infection detection time of \system{} is far better than all other approaches. This is due to the simple heuristic algorithm of comparing only three features from each packet. \system{} is able to detect the very first packet generated by malware, and it requires only 1.6 ms on average to do so. In comparison, N-BaIoT needs 174 ms and D\"IoT requires 256 ms for detection. The heavy machine learning models used in N-BaIoT (Autoencoder) and D\"IoT (Gated Recurrent Unit) cause these longer detection times. The detection time of VRASED depends on the frequency of remote attestation, i.e., how soon it is performed after infection, plus the time required to complete attestation. Additionally, the attestation time is proportional to memory size since VRASED calculates HMAC over the entire memory as part of the operation. It takes 450 ms to complete attestation for 4KB memory. This attestation time is significantly longer compared to \system{}'s. \system{}'s FPR is slightly higher compared to the other approaches. However, this FPR for off-the-shelf devices is overestimated since it is calculated without considering the feedback loop. The same packet is counted as false positive many times. For example, the packet with external address \textit{www.google.com}, packet direction \textit{Server-to-Device}, and length \textit{409} is counted 13 times in the calculation of the FPR of the Google Home Mini device from BehavIoT dataset. With the feedback loop, the packet would only be considered suspicious once. 
\begin{table}[t!]
	\scriptsize
	\centering
        \caption{Performance comparison of \system{} with prior approaches.}
	\begin{tabularx}{\linewidth}
{r p{15mm} r r r}
  \toprule
  \textbf{Approach} & \textbf{\# Required Features} & \textbf{TPR} & \textbf{FPR} & \textbf{Detection Time (ms)} \\
  \midrule
  \textbf{\system{}} & \textbf{3} & \textbf{100.0\%} & \textbf{0.78\%} & \textbf{1.6}\\
  N-BaIoT \cite{meidan2018n} & 114 & 100.0\% & 0.007\% & 174 \\
  D\"IoT \cite{nguyen2019diot} & 7 & 95.6\% & 0\% & 256 \\
  VRASED \cite{vrasedp} & N/A & 100.0\% & 0\% & attestation freq. + 450\\
  \bottomrule
\end{tabularx}
\vspace{-1em}
\label{tab:performance comparison}
\end{table}

\subsection{Energy Savings}
\label{subsec: energy-savings}
In this section, we attempt to quantify the energy saved by the feedback loop of \system{} over plain periodic remote attestation. For this purpose, we consider two IoT devices: (1) DCS-930L Security Camera \cite{DCS-930L}, a higher-end device that was compromised by Mirai \cite{nguyen2019diot}; and (2) TI MPS430, a bare metal IoT device, on top of which VRASED is implemented. Table~\ref{tab:attestation energy} shows the summary of the energy consumption rates for periodic \ra{} on these two devices in a year.

DCS-930L camera is not capable of \ra{}. However, if it was, we assume that its \ra{} time would be similar to the emulated RPi bulb since both have comparable operating systems and processing powers. The power consumption of the DCS-930L camera is 1.84 W~\cite{DCS-930L-specs}, whereas the \ra{} time is considered to be 1.6 ms. This results in 0.49 mWh energy consumption per \ra{}. If \ra{} is performed once every 5 minutes to detect an attack within a reasonable time frame, the total energy consumption to perform \ra{} over a year becomes 5,157.9 mWh. Energy consumption for ``somewhat instantaneous'' detection (\ie{} within 1 minute of infection) is even higher, namely 25,789.4 mWh or $\sim$26 Wh. \textit{In other words, the camera would need at least $\sim$14$\times$ the amount energy that it needs to operate in one hour, just to perform \ra{} in a year without any guarantee of \textit{actually} detecting the malware.}

TI MSP430 consumes 1.2 mWh power in active mode \cite{TI-MSP430-specs}, and the attestation time of VRASED is 450 ms. This results in 0.009 mWh energy per attestation, 946.1 mWh over the year when the attestation period is 5 minutes and 4,730.4 mWh for an attestation period of 1 minute. This is a critical energy burden for an ultra-low power device that requires only 400 $\mu$A current during the active period. 

Energy consumption of periodic attestation can be drastically reduced by adopting \system{} since \system{} triggers attestation \textit{only} when there is a suspicious packet because of the rare events: actual malware infections or firmware updates.
\begin{table}[t!]
	\scriptsize
	\centering
        \caption{Comparison on total yearly energy consumption of periodic \ra{} (Without \system{}) for various intervals.}
	\begin{tabularx}{\linewidth}
{r r r r r r}
  \toprule
  \textbf{Device Name} & \multicolumn{5}{c}{\textbf{Energy Consumption (mWh) for Attestation Intervals}} \\
  \cline{2-6}
  & \textbf{1 hour} & \textbf{30 mins} & \textbf{10 mins} & \textbf{5 mins} & \textbf{1 min}\\
  \midrule
  DCS-930L Cam. & 423.9 & 859.6 & 2,578.9 & 5,157.9 & 25,789.4 \\
  TI MPS430 & 77.8 & 157.7 & 473.0 & 946.1 & 4,730.4\\
  \bottomrule
\end{tabularx}
\vspace{-1em}
\label{tab:attestation energy}
\end{table}

%% file: content/discussion.tex
\section{Discussion and Limitations} \label{sec:discussion}

\subsection{\system{}'s Scope}
\parheading{Closing the Loop.} 
We acknowledge that \system{} does not quite ``close the loop" on malware detection. The missing component is the feedback from scheduled \ra to \ta. Ideally, whenever scheduled/periodic  (i.e., not triggered by \ta) \ra{} detects malware, \ta heuristics would be adjusted based on recent traffic logs that must contain infection-relevant traffic. Otherwise, the device must have been infected via direct/physical
means. This is part of future work.

\parheading{Malware without Network Traffic}
\system cannot detect malware that sits on a device silently without generating any network traffic. 
However, if such malware is transmitted to the device through the network, then \system catches it during transmission. 
Additionally, certain types of Mirai-variant malware have the capability to remain dormant for a specific duration before becoming active to initiate a DoS attack. 
In such scenarios, the \system can detect the malware when it becomes active and starts generating network traffic.
\subsection{Attestation TCB} \label{subsec:attestationtcb}
In \system Attester is implemented by extending the operating system (OS). This increases the TCB size by including the OS in the TCB and makes it vulnerable. 
To mitigate this, a TPM~\cite{tpm} can be integrated and a secure
boot process can be configured to verify the TCB binary’s
validity. This ensures that the OS always boots in the expected
state. However, advanced attacks such as privilege escalation
and runtime control-flow attacks can still compromise the
TCB, which is out of scope for this paper.

\subsection{Compatibility with Other Attestation Techniques}
There are more secure ways of implementing attestation in IoT devices besides the OS-based Attester module. 
Notably, two options would be: (1) using off-the-shelf ARM TrustZone, or (2) using theoretical Root-of-Trusts (RoTs) such as SMART~\cite{smart} and Sancus~\cite{sancus}.

ARM TrustZone-M~\cite{ARM-TrustZone-M} provides secure isolation between normal and secure regions of memory for devices with ARM Cortex-M23/33/55 microcontroller units (MCUs). Installing the attestation TCB in the secure region prevents illegal access or modification by programs in the normal region. Further, the TCB can verify if the device is running expected user binary by accessing and measuring the normal region when an attestation request is received. We implemented such a technique on NXP LPC55S69-EVK \cite{nxpboard} board with ARM Cortex-M33. However, we could not test and evaluate the prototype due to lack of real-world malware on low-end devices. Higher-end processors such as ARM Cortex-A32/72 are equipped with TrustZone-A  ~\cite{trustzone}, Trusted Execution Environment (TEE) extensions, which can access the physical memory of normal regions, but it is difficult to implement an attestation scheme that can measure user processes. This is because the normal region's separate Memory Management Unit (MMU) virtualizes user processes for the normal OS. Creating an attestation mechanism for TrustZone-A without trusting the normal OS is a non-trivial task. For ultra-low-power MCUs like TI MSP430~\cite{msp_memory_specs} and AVR ATMega8~\cite{atmel_specs}, several RoTs~\cite{smart,sancus,vrasedp,trustlite} propose attestation techniques based on custom hardware extensions. Since \system{} only relies on Attester as a black box, it can seamlessly integrate with these RoTs without modifications.

\subsection{Physical Attacks}
There are two types of physical attacks: non-invasive and invasive. Both are considered out of the scope of this paper.

Non-invasive attacks involve physical reprogramming of device software using direct/wired interfaces, such as USB/UART, SPI, or I2C. These attacks are challenging to detect, particularly when the attestation TCB becomes compromised or when the measurement values are maliciously updated to reflect the corrupted state of the applications being attested. 
However, well-known security mechanisms, such as secure boot, can mitigate these attacks.

Invasive attacks, on the other hand, involve attempting physical damage to the device, such as inducing hardware faults, tampering with hardware components, or exploiting physical side channels to extract secrets. Protection against invasive attacks can be achieved through standard tamper-resistant techniques~\cite{ravi2004tamper}.

\subsection{Mitigation of DoS attacks}
There are several ways an adversary can launch a DoS attack on an IoT device.
Recall from Section \ref{subsec:advesarymodel} that an adversary can overwhelm an IoT device by sending a large volume of packets, or it can utilize the infected devices to flood the network with excessive traffic, affecting other devices in the same network.
These attacks can be detected when the gateway observes abnormal traffic to/from an IoT device, triggering attestation.
Subsequently, since an IoT device is unresponsive during the attack, the attestation request will be disregarded, and after a timeout, \system will raise a failure.

Next, the adversary can also trick the gateway by sending abnormal packets to an IoT device and forcing the gateway to perform attestation repeatedly. 
Repeated attestation not only consumes power but can also potentially degrade the functionality of an IoT device.
Although such attacks are out of the scope of \system,  we note that they can be mitigated by implementing rate-limiting to perform attestation only a fixed number of times within a certain interval.

Furthermore, if the adversary has prior knowledge of the IoT device profile, it can launch a DoS attack by sending a burst of packets that conform to the profile.
For instance, the adversary can construct malicious packets with the same packet length and forged source address to match one of the profile entries.
In such cases, Monitor would allow the traffic to pass, resulting in a successful attack.  
Such attacks are also out-of-scope.
Note that detecting such traffic would require stateful profile monitoring to check the timing interval between consecutive packets. 
However, implementing such monitoring is computational and storage intensive.
Considering the general IoT traffic observed in our evaluation, we believe the benefits of deploying such resource-intensive measures are not substantial enough to justify their implementation.

\subsection{Software Updates} 
There can be two different types of software updates: (1) the server's software changes resulting in different external address and/or packet lengths. However, the software state of the IoT device remains unchanged; or (2) the IoT device's software state changes.
Similar to a new command that generates benign traffic (see Section~\ref{subsec:effectiveness-feedback-loop}), software updates of the first type can be handled by the feedback loop. 
When a software update for the IoT device occurs, it triggers Monitor with a new packet. Monitor then calls Attester on the IoT device to perform attestation, which does not detect any change in the software state. Finally, this new packet is included as part of the IoT device's profile. It can also be handled manually by providing the gateway with the updated profile information. 
However, software updates of the second type cannot be incorporated automatically by \system{}. This is because when called by Monitor, Attester sees that the software state does not match the reference value. As a result, the legitimate new packet is considered malicious. To solve this problem, the server must provide the updated reference value to the IoT device and gateway in this case.

\subsection{IoT Devices in Ad Hoc Network}
IoT devices in ad hoc mode can exchange packets without going through the gateway, making it impossible for Profiler or Monitor to capture and process them. Such devices are out of the scope of \system{}. However, we assume that the owner of the IoT device is trusted. Thus, in other words, the IoT device is configured to operate in infrastructure mode that allows all packets to go through the gateway.  

%% file: content/conclusion.tex
\section{Conclusion}\label{sec:conclusion}
We present \system{}, a system that combines attestation and network traffic analysis to detect malware in IoT devices. \system{} stores benign traffic patterns from IoT devices that are usually limited and predictable. On the other hand, devices with malware infection exhibits network traffic patterns that are likely divergent from the usual traffic pattern. When unusual patterns are detected, \system{} performs \ra{} on the IoT device that produces suspicious traffic pattern. If malware infection is confirmed, \system{} will report this incident; otherwise, it will whitelist the traffic pattern as benign traffic. 
Our evaluation shows that \system{} is both effective and efficient in profiling and monitoring uninfected and infected devices. It achieves 100\% TPR with low false positive rate (FPR) in most cases with low runtime (\ie{} $\sim$1.6 ms traffic classification time) and storage overheads. Finally, it outperforms the other approaches in terms of efficiency and effectiveness.

%% file: content/appendix.tex
\renewcommand{\thetable}{A\arabic{table}}
\setcounter{table}{0}

\section*{Appendix A: Complete Device Profile}
\label{appendix: profile}
This appendix provides the complete device profile of an off-the-shelf device, Sensi Thermostat in Table \ref{tab: sensi thermostat profile detailed}. The partial profile of this device is shown in Figure \ref{fig: profile example} of Section \ref{sec:design}.
    \begin{table*}[t!]
	\centering
        \caption{Complete Device Profile of Sensi Thermostat.}
        \footnotesize
\begin{tabularx}{\textwidth}{p{3cm} p{4cm} p{4cm} p{2cm} p{2cm}} 
  \toprule
    \textbf{DEVICE MAC} & \textbf{PACKET DIRECTION} & \textbf{EXTERNAL ADDRESS} &  \textbf{PACKET LENGTH} & \textbf{Device Name}\\ \hline
34:6f:92:1d:ba:50 &	USER\_TO\_DEVICE   &   	192.168.4.1	&			62 &	Sensi Thermostat \\ \hline
34:6f:92:1d:ba:50 &	DEVICE\_TO\_SERVER &   	224.0.0.251	&			46 &	Sensi Thermostat \\ \hline
34:6f:92:1d:ba:50 &	DEVICE\_TO\_SERVER &   	icda.sensicomfort.com &	58 &	Sensi Thermostat \\ \hline
34:6f:92:1d:ba:50 &	USER\_TO\_DEVICE   &   	192.168.4.1			  &	86 &	Sensi Thermostat \\ \hline
34:6f:92:1d:ba:50 &	SERVER\_TO\_DEVICE &   	icda.sensicomfort.com &	58 &	Sensi Thermostat \\ \hline
34:6f:92:1d:ba:50 &	DEVICE\_TO\_SERVER &   	icda.sensicomfort.com &	54 &	Sensi Thermostat \\ \hline
34:6f:92:1d:ba:50 &	DEVICE\_TO\_SERVER &   	icda.sensicomfort.com &	104 &	Sensi Thermostat \\ \hline
34:6f:92:1d:ba:50 &	SERVER\_TO\_DEVICE &   	icda.sensicomfort.com &	54 &   	Sensi Thermostat \\ \hline
34:6f:92:1d:ba:50 &	SERVER\_TO\_DEVICE &   	icda.sensicomfort.com &	144 &   Sensi Thermostat \\ \hline
34:6f:92:1d:ba:50 &	DEVICE\_TO\_SERVER &   	icda.sensicomfort.com &	255 &   Sensi Thermostat \\ \hline
34:6f:92:1d:ba:50 &	DEVICE\_TO\_SERVER &	icda.sensicomfort.com &	129	 & Sensi Thermostat \\ \hline
34:6f:92:1d:ba:50 &	SERVER\_TO\_DEVICE &	icda.sensicomfort.com &	129	 & Sensi Thermostat \\ \hline
34:6f:92:1d:ba:50 &	DEVICE\_TO\_SERVER &	icda.sensicomfort.com &	139	 & Sensi Thermostat \\ \hline
34:6f:92:1d:ba:50 &	SERVER\_TO\_DEVICE &	icda.sensicomfort.com &	139	 & Sensi Thermostat \\ \hline
34:6f:92:1d:ba:50 &	DEVICE\_TO\_SERVER &	icda.sensicomfort.com &	539	 & Sensi Thermostat \\ \hline
34:6f:92:1d:ba:50 &	SERVER\_TO\_DEVICE &	icda.sensicomfort.com &	155	 & Sensi Thermostat \\ \hline
34:6f:92:1d:ba:50 &	DEVICE\_TO\_SERVER &	icda.sensicomfort.com &	427	 & Sensi Thermostat \\ \hline
34:6f:92:1d:ba:50 &	DEVICE\_TO\_SERVER &	icda.sensicomfort.com &	619	 & Sensi Thermostat \\ \hline
34:6f:92:1d:ba:50 &	SERVER\_TO\_DEVICE &	icda.sensicomfort.com &	299	 & Sensi Thermostat \\ \hline
34:6f:92:1d:ba:50 &	DEVICE\_TO\_SERVER &	icda.sensicomfort.com &	491	 & Sensi Thermostat \\ \hline
34:6f:92:1d:ba:50 &	DEVICE\_TO\_SERVER &	icda.sensicomfort.com &	299	 & Sensi Thermostat \\ \hline
34:6f:92:1d:ba:50 &	DEVICE\_TO\_SERVER &	icda.sensicomfort.com &	267	 & Sensi Thermostat \\ \hline
34:6f:92:1d:ba:50 &	DEVICE\_TO\_SERVER &	icda.sensicomfort.com &	443	 & Sensi Thermostat \\ \hline
34:6f:92:1d:ba:50 &	DEVICE\_TO\_SERVER &	icda.sensicomfort.com &	219	 & Sensi Thermostat \\ \hline
34:6f:92:1d:ba:50 &	DEVICE\_TO\_SERVER &	icda.sensicomfort.com &	107	 & Sensi Thermostat \\ \hline
34:6f:92:1d:ba:50 &	SERVER\_TO\_DEVICE &	icda.sensicomfort.com &	107	 & Sensi Thermostat \\ \hline
34:6f:92:1d:ba:50 &	DEVICE\_TO\_SERVER &	icda.sensicomfort.com &	635	 & Sensi Thermostat \\ \hline
34:6f:92:1d:ba:50 &	SERVER\_TO\_DEVICE &	icda.sensicomfort.com &	619	 & Sensi Thermostat \\ \hline
34:6f:92:1d:ba:50 &	DEVICE\_TO\_SERVER &	icda.sensicomfort.com &	283	 & Sensi Thermostat \\ \hline
34:6f:92:1d:ba:50 &	SERVER\_TO\_DEVICE &	icda.sensicomfort.com &	459	 & Sensi Thermostat \\ \hline
34:6f:92:1d:ba:50 &	DEVICE\_TO\_SERVER &	icda.sensicomfort.com &	155	 & Sensi Thermostat \\ \hline
34:6f:92:1d:ba:50 &	SERVER\_TO\_DEVICE &	icda.sensicomfort.com &	1450 & Sensi Thermostat \\ \hline
34:6f:92:1d:ba:50 &	SERVER\_TO\_DEVICE &	icda.sensicomfort.com &	172	 & Sensi Thermostat \\ \hline
34:6f:92:1d:ba:50 &	SERVER\_TO\_DEVICE &	icda.sensicomfort.com &	864	 & Sensi Thermostat \\ \hline
34:6f:92:1d:ba:50 &	SERVER\_TO\_DEVICE &	icda.sensicomfort.com &	512	 & Sensi Thermostat \\ \hline
34:6f:92:1d:ba:50 &	SERVER\_TO\_DEVICE &	icda.sensicomfort.com &	917	 & Sensi Thermostat \\ \hline
34:6f:92:1d:ba:50 &	DEVICE\_TO\_SERVER &	icda.sensicomfort.com &	61	 & Sensi Thermostat \\ \hline
34:6f:92:1d:ba:50 &	DEVICE\_TO\_SERVER &	192.168.47.2		  &	54	 & Sensi Thermostat \\ \hline
34:6f:92:1d:ba:50 &	DEVICE\_TO\_SERVER &	icda.sensicomfort.com &	330	 & Sensi Thermostat \\ \hline
34:6f:92:1d:ba:50 &	SERVER\_TO\_DEVICE &	icda.sensicomfort.com &	61	 & Sensi Thermostat \\ \hline
34:6f:92:1d:ba:50 &	DEVICE\_TO\_SERVER &	icda.sensicomfort.com &	475	 & Sensi Thermostat \\ \hline
34:6f:92:1d:ba:50 &	DEVICE\_TO\_SERVER &	icda.sensicomfort.com &	411	 & Sensi Thermostat \\ \hline
34:6f:92:1d:ba:50 &	DEVICE\_TO\_SERVER &	icda.sensicomfort.com &	171	 & Sensi Thermostat \\ \hline
34:6f:92:1d:ba:50 &	DEVICE\_TO\_SERVER &	icda.sensicomfort.com &	571	 & Sensi Thermostat \\ \hline
34:6f:92:1d:ba:50 &	SERVER\_TO\_DEVICE &	icda.sensicomfort.com &	118  & Sensi Thermostat \\ \hline
34:6f:92:1d:ba:50 &	SERVER\_TO\_DEVICE &	icda.sensicomfort.com &	224	 & Sensi Thermostat \\ \hline
34:6f:92:1d:ba:50 &	SERVER\_TO\_DEVICE &	icda.sensicomfort.com &	384	 & Sensi Thermostat \\ \hline
34:6f:92:1d:ba:50 &	SERVER\_TO\_DEVICE &	icda.sensicomfort.com &	192	 & Sensi Thermostat \\ \hline
34:6f:92:1d:ba:50 &	SERVER\_TO\_DEVICE &	icda.sensicomfort.com &	315	 & Sensi Thermostat \\ \hline
34:6f:92:1d:ba:50 &	SERVER\_TO\_DEVICE &	icda.sensicomfort.com &	789	 & Sensi Thermostat \\ \hline
34:6f:92:1d:ba:50 &	DEVICE\_TO\_SERVER &	icda.sensicomfort.com &	315	 & Sensi Thermostat \\ \hline
34:6f:92:1d:ba:50 &	DEVICE\_TO\_SERVER &	icda.sensicomfort.com &	347	 & Sensi Thermostat \\ 
\bottomrule
\end{tabularx}
\vspace{-1em}
    \label{tab: sensi thermostat profile detailed}
\end{table*}

\renewcommand{\thefigure}{B\arabic{figure}}
\setcounter{figure}{0}

\section*{Appendix B: Difference Between Malware Packets \& IoT Device Packets}
\label{appendix: packet  difference}
In this appendix we briefly describe the difference between malware packets and IoT device packets that we observed in the course of developing \system.

\begin{figure}[!ht]
    \centering
    \includegraphics[width=0.6\linewidth]{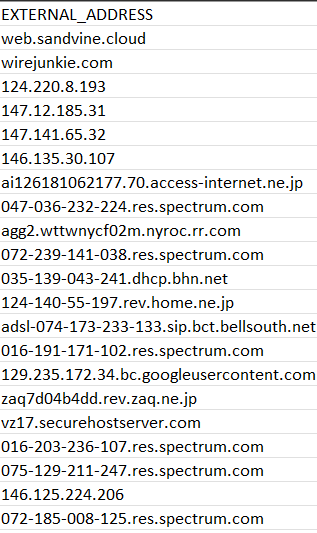}
    \caption{Some of the external addresses used by malware.} 
    \label{fig: malware address}
\end{figure}

\parheading{External Addresses.}
Packets generated by malware binaries have completely different external addresses from external addresses of packets generated by the observed IoT devices. Most malware programs try to connect with their CCC using static IP addresses. \system is sometimes able to map these IP addresses to hostnames through reverse DNS lookup. Interestingly, we observe that none of those hostnames indicates a service hosted in AWS cloud, whereas most hostnames found through reverse DNS lookup for IoT devices belong to AWS cloud. However, few malware programs use hostnames resolved through DNS lookup. Figure~\ref{fig: malware address} shows a partial list of external addresses found for malware packets.

\begin{figure}[!ht]
    \centering
    \includegraphics[width=\linewidth]{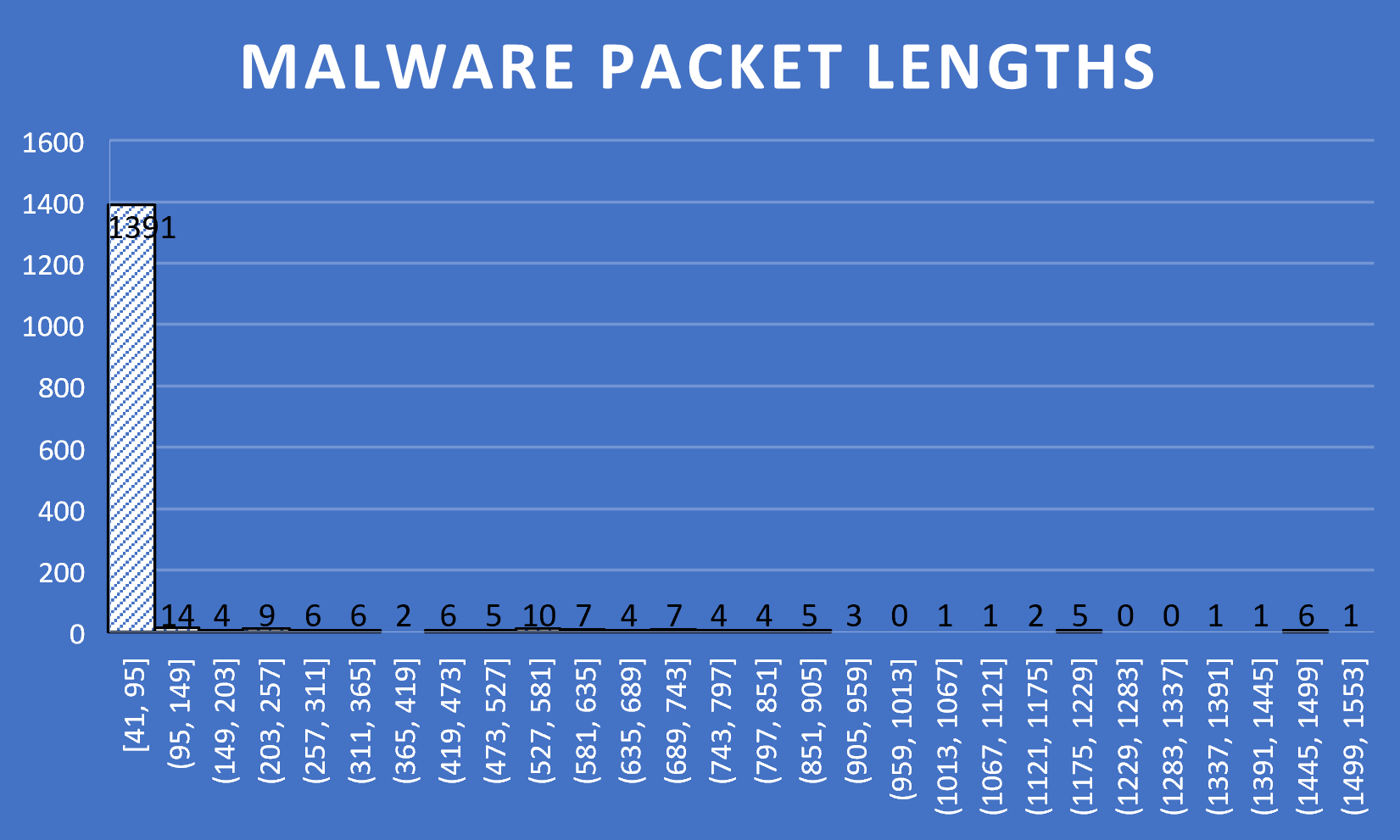}
    \caption{Length Distribution of Malware Packets.} 
    \label{fig: malware packet length}
\end{figure}
\begin{figure}[!ht]
    \centering
    \includegraphics[width=\linewidth]{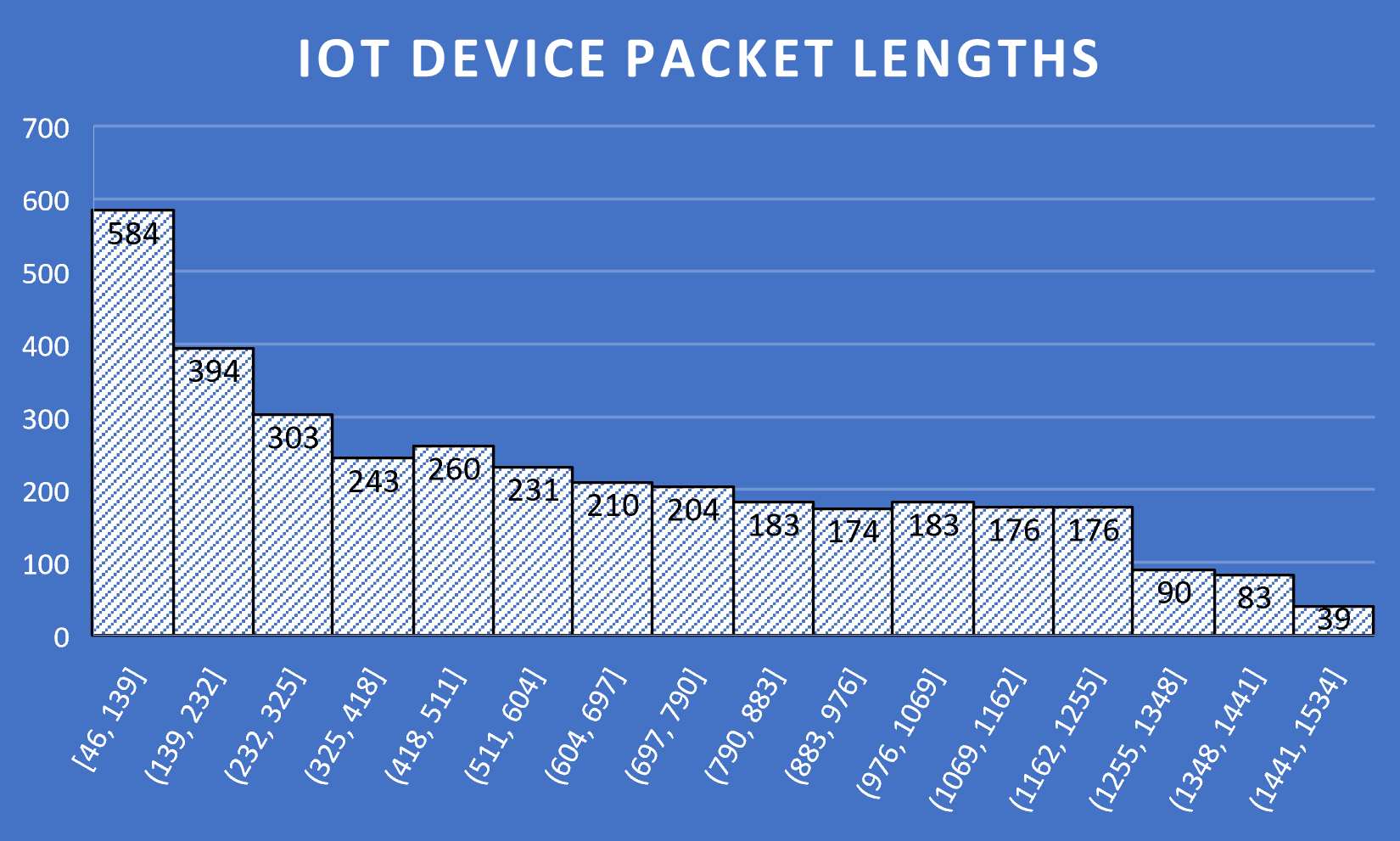}
    \caption{Length Distribution of IoT Device Packets.} 
    \label{fig: device packet length}
\end{figure}

\parheading{Packet Lengths.}
The length distribution of malware packets is also drastically different from the length distribution of IoT device packets. This is because observed malware packets are mostly TCP SYN, ACK, and RST packets. Figure \ref{fig: malware packet length} and Figure \ref{fig: device packet length} show the length distributions for malware packets and device packets respectively. 